\documentclass[prd,aps,preprintnumbers, showpacs, nofootinbib,superscriptaddress,
notitlepage
]{revtex4-1}
\usepackage{epsfig}
\usepackage{bm}
\usepackage{amssymb}
\usepackage{amsmath}
\usepackage{color}
\usepackage{subfigure}

\newcommand{\be}{\begin{equation}}
\newcommand{\ee}{\end{equation}}
\newcommand{\bea}{\begin{eqnarray}}
\newcommand{\eea}{\end{eqnarray}}

\newcommand{\beq}{\begin{eqnarray}}
\newcommand{\eeq}{\end{eqnarray}}

\begin{document}

\title{Elliptic flow difference of charged pions in heavy-ion collisions}

\preprint{YITP-15-57}

\author{Yoshitaka Hatta}
\affiliation{Yukawa Institute for Theoretical Physics, Kyoto University, Kyoto 606-8502, Japan}

\author{Akihiko Monnai}
\affiliation{RIKEN BNL Research Center, Brookhaven National Laboratory, Upton 11973 NY, USA}

\author{Bo-Wen Xiao}
\affiliation{Key Laboratory of Quark and Lepton Physics (MOE) and Institute
of Particle Physics, Central China Normal University, Wuhan 430079, China}

\date{\today}

\begin{abstract}

Recently, the STAR collaboration at RHIC has presented experimental evidence for the correlation between the elliptic flow difference of charged pions and charge asymmetry as a possible signal of the chiral magnetic wave. We demonstrate that the STAR results can be understood within the standard viscous hydrodynamics.

\end{abstract}

\maketitle


Recently, the STAR collaboration at RHIC has measured  the difference in the elliptic flow parameter $v_2$ between $\pi^+$ and $\pi^-$ on an event-by-event basis, and found a linear dependence on the charge asymmetry of the collision system $A_{ch}\equiv \frac{N_+ -N_-}{N_+ + N_-}$ \cite{Adamczyk:2015eqo}
\beq
\Delta v_2^{\pi}\equiv v_2(\pi^-) -v_2(\pi^+) = \Delta v_2^\pi(\textrm{base})+ r A_{ch}\,,
 \label{dif}
\eeq
 where $N_{\pm}$ is the multiplicity of positively (negatively) charged particles. This quantity has attracted much interest over the past several years as a possible signal of the so-called chiral magnetic wave (CMW)  \cite{Burnier:2011bf}. It has been argued that a strong magnetic field created in off-center heavy-ion collisions together with the chiral anomaly induce a quadrupole deformation of electric charges, resulting in the elliptic flow difference (\ref{dif}). The main predictions of the CMW are that $\Delta v_2^{\pi}$ depends linearly on $A_{ch}$, and the slope parameter $r$ is positive. Moreover, $r$ has a characteristic peak as a function of the centrality. The STAR data are qualitatively similar to these expectations.

Of course, one has to thoroughly examine all the other `non-exotic' mechanisms that can  contribute to the difference  (\ref{dif}) and subtract their contributions before finally claiming the discovery of the CMW in heavy-ion collisions.  Some attempts in this direction have been  made in \cite{Bzdak:2013yla,Campbell:2013ika}, but these alternative scenarios are already disfavored by the data \cite{Adamczyk:2015eqo} (see, however, \cite{Hongo:2013cqa}).  In this paper, we raise the possibility that the STAR data  can actually be understood within the standard viscous hydrodynamics.

In a previous paper \cite{Hatta:2015era}, we have analytically computed the difference $\Delta v_2^{\pi}$ for the anisotropic Gubser flow \cite{Gubser:2010ui} coupled with conserved currents and found that it is proportional to the shear viscosity $\eta$ and the isospin chemical potential $\mu_I$. The result is, neglecting numerically small corrections,
 \beq
\frac{\Delta v_2^{\pi}}{v^{\pi, ideal}_2} \approx \frac{-\mu_I}{T}\frac{27}{80} K\,, \label{our}
\eeq
where $K\propto \eta/s$  is the Knudsen number ($s$ is the entropy density) and $T$ is the freezeout temperature.
(The sign convention of $\Delta v_2$ here is different from \cite{Hatta:2015era}.) In heavy-ion collisions, the mean value of $\mu_I$ is slightly negative because the colliding nuclei are neutron-rich. In place of $\mu_I$, we may alternatively use the charge chemical potential $\mu_Q$. Numerically, $\mu_I$ are $\mu_Q$ are close (cf, Ref.~\cite{Bazavov:2012vg}), and we expect that the following discussion will be qualitatively similar in the two cases.

In order to establish the connection between (\ref{dif}) and (\ref{our}),
we
classify events according to the value of $A_{ch}$ and assign effective freezeout parameters in each bin of $A_{ch}$.\footnote{To assign freezeout parameters for certain subevents is a common practice in heavy-ion collisions. For instance, $T$ and $\mu$ are often plotted as a function of centrality at fixed energy. Experimentally, centrality is determined by multiplicity, and the idea here is that events with the same multiplicity are regarded as a statistical ensemble and chemical potentials can be independently assigned for this ensemble. Here we do the same, using $A_{ch}$ bins instead of multiplicity bins. In the STAR measurements, the number of events in each bin of  $A_{ch}$ at fixed centrality is typically ${\mathcal O}(10^5)$. }   This can be done by using the statistical model of hadrons in which  the multiplicity $N_i$ of hadron species $i$ is computed as \cite{Wheaton:2004qb}
\beq
N_i=\frac{g_i V}{2\pi^2}\sum_{k=1}^\infty(\mp 1)^{k+1}\frac{m_i^2 T}{k}K_2\left(\frac{km_i}{T}\right)
\exp\left(k\frac{B_i\mu_B +I_i \mu_I + S_i\mu_S}{T}\right)\,, \label{stat}
\eeq
 where  $V$ is the volume, $m_i$ is the mass and $g_i$ is the degeneracy factor.\footnote{The temperature $T$ here is the chemical freezeout temperature which in general differs from the kinetic freezeout temperature in (\ref{our}). Here we assume that the two temperatures are equal or close to each other, following the early freezeout model \cite{Hatta:2015era} in which the formula (\ref{our}) was derived. Note that in the same model the ratio $\mu_I/T$ in (\ref{our}) is approximately constant during the time evolution.} The sign $\mp 1$ corresponds to fermions/bosons. $B,I,S$ are the baryon, isospin and strangeness quantum numbers, respectively, and $\mu_{B,I,S}$ are the corresponding chemical potentials. The asymmetry $A_{ch}$ can be evaluated by summing over  all charged hadrons $N_\pm=\sum_i N_\pm^i$ whose masses are below 2 GeV.\footnote{ In practice, we include the $k=1,2,3$ terms in the sum (\ref{stat}) for pions, $k=1,2$ terms  for kaons and only the $k=1$ term for all the other hadrons. We have checked that other higher $k$ terms are negligible.} This determines a function $A_{ch}(\mu_{B},\mu_I,\mu_S,T)$ which is to a very good approximation  linear\footnote{ At low  energy where $\mu_B\gtrsim 200\,$MeV, we start to see weak deviations from the linear fit (\ref{lin}) which however do not   obstruct the determination of $r$.} in $\mu$'s,
 \beq
 A_{ch} \approx c(T) \mu_B + c'(T)\mu_I + c''(T) \mu_S\,, \label{lin}
 \eeq
  with all the coefficients (`susceptibilities') positive $c'>c''>c>0$.\footnote{ At $T=159\ $MeV, and when the linear approximation is valid for $A_{ch}$, we find $c=0.383$, $c'=5.45$ and $c''=1.45$ in units of $\textrm{GeV}^{-1}$.}  Naively,   there seems to be a sign mismatch if one uses (\ref{lin}) to rewrite (\ref{our}) in the form (\ref{dif}) because $r>0$ and $c'>0$. However, one should take into account the fact that in heavy-ion collisions $\mu_{B},\mu_I$ and $\mu_S$ are not independent of each other.
The ratios $\mu_I/\mu_B$ and $\mu_S/\mu_B$ are more or less universal as they are determined from the quantum numbers of the colliding nuclei, namely, isospin asymmetry and strangeness-free conditions \cite{Stachel:2013zma,Bazavov:2012vg}. These conditions can be and have been used at different energies, different centrality bins and rapidity windows (i.e., different types of subevents). We thus assume that these ratios are fixed.
 More precisely, in practice, we consider the following two parameterizations which we extracted from the result of the statistical model fits of experimental data from the SPS to RHIC, $7.6\, \mbox{GeV}\le \sqrt{s_{NN}}\le 200$ GeV \cite{Stachel:2013zma}
   \begin{eqnarray}
  \mu_I =-0.0308\mu_B+2.77 \cdot 10^{-8}\mu_B^3\,,  \qquad
  \mu_S = 0.249\mu_B -1.09\cdot 10^{-7}\mu_B^3\,, \label{lat}
 \end{eqnarray}
  \begin{eqnarray}
  \mu_I = -0.293-0.0264\mu_B\,,  \qquad
  \mu_S = 1.032 +0.232\mu_B\,, \label{inter}
 \end{eqnarray}
where $\mu$'s are in units of MeV. The first choice (\ref{lat}) which includes cubic terms is motivated by a recent lattice study \cite{Bazavov:2012vg}. The second choice allows for possible intercepts in the limit $\mu_B\to 0$.  Actually, there is some arbitrariness in defining  $\mu_I$  when $\mu_B<0$. Yet, keeping the same ratios $\mu_I/\mu_B$ and $\mu_S/\mu_B$ is natural from the viewpoint of the charge conjugation symmetry.

\begin{figure}[tbp]
  \includegraphics[width=100mm]{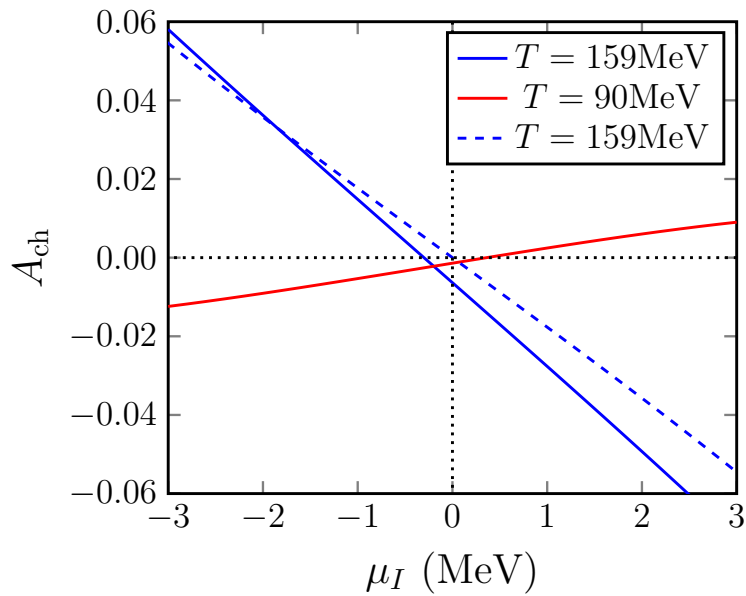} 
 \caption{Charge asymmetry $A_{ch}$ as a function of the isospin chemical potential $\mu_I$. Blue solid line: $T=159$ MeV with the parameterization  (\ref{inter}). Blue dashed line: $T=159$ MeV with (\ref{lat}). Red line: $T=90$ MeV with  (\ref{inter}). \label{fig1}}
\end{figure}

With the above parameterizations, $A_{ch}$ becomes a function of $\mu_I$ and $T$. This is shown in Fig.~\ref{fig1} at $T=159$ MeV relevant to $\sqrt{s_{NN}}=200$ GeV at RHIC \cite{Stachel:2013zma}.
 We observe that, somewhat counterintuitively,  $A_{ch}$ is a decreasing function of $\mu_I$. Actually, the slope is  sensitive to the temperature. As also shown in  Fig.~\ref{fig1}, if we artificially lower the temperature to, say, $T=90$ MeV, $A_{ch}$ becomes an increasing function of $\mu_I$. This indicates that, at high temperature, the charge asymmetry is not dominated by pions.

\begin{figure}[tbp]
  \includegraphics[height=6.7cm]{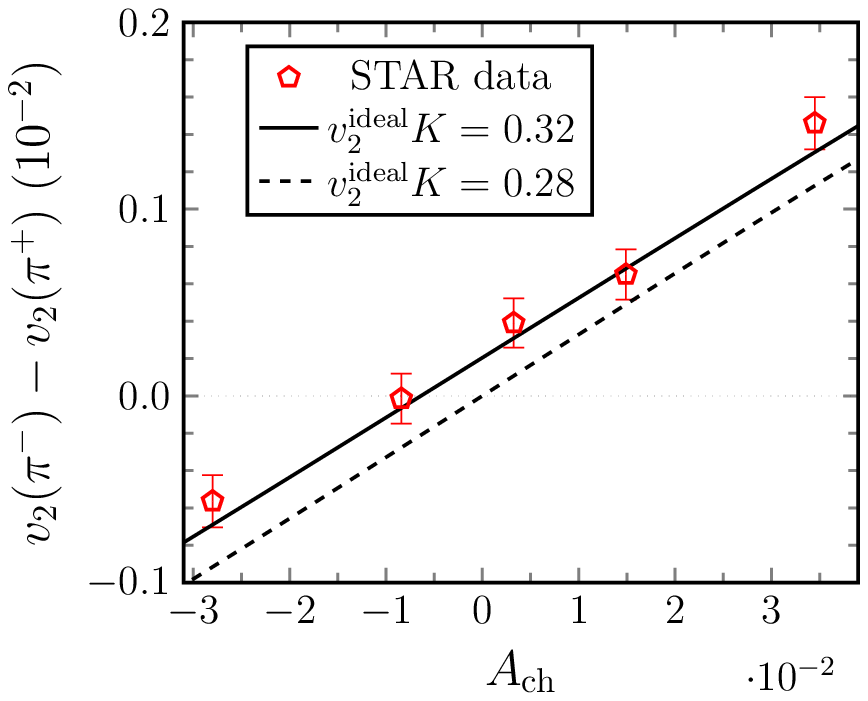}
  \includegraphics[height=6.7cm]{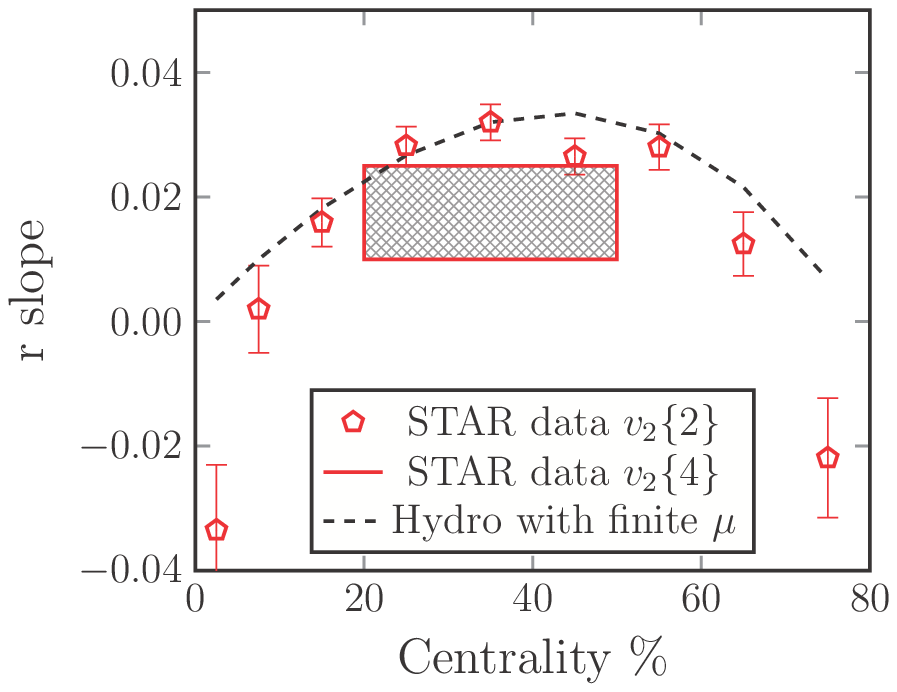}
 \caption{Left figure: $\Delta v_2^{\pi}$ versus $A_{ch}$ with $T=159$ MeV at $\sqrt{s_{NN}} =200 \,\textrm{GeV}$, 30-40\% centrality. The solid and dashed lines correspond to (\ref{inter}) and (\ref{lat}), respectively. Right figure: The slope $r$ as a function of centrality at $\sqrt{s_{NN}} =200\, \textrm{GeV}$. The data point $r=0.032$ at 35\% centrality is the input for the theory curve. The STAR data labelled with $v_2\{2\}$ and $v_2\{4\}$ are obtained by using the two-particle and four-particle cumulant methods, respectively. Only statistical error bars are shown in the above plots.\label{fig2}}
\end{figure}

We thus find that (\ref{dif}) and (\ref{our}) are consistent including the sign of $r$. The magnitude of $r$, on the other hand, depends both on the centrality and  the collision energy as observed by the STAR collaboration \cite{Adamczyk:2015eqo}. We  confront these data with our theoretical results in \cite{Hatta:2015era}. For this purpose, we first fix the normalization of $v_2^{ideal}K$ in (\ref{our}) by fitting  the slope $r=0.032$ measured in the 30-40\% centrality range at $\sqrt{s_{NN}}=200$ GeV. We find $v_2^{\pi,ideal}K = 0.28$ and $v_2^{\pi,ideal}K = 0.32$ in the two cases (\ref{lat}) and (\ref{inter}), respectively, see Fig.~\ref{fig2}\,(left). Note that in this fit the `intercept' $\Delta v_2^\pi(\mbox{base})$ in (\ref{dif}), which was experimentally found to be nonzero and positive, is automatically reproduced in the case (\ref{inter}). See Ref.~\cite{Hongo:2013cqa,Stephanov:2013tga} for CMW-based derivations of $\Delta v_2(\mbox{base})$.

In the following, we shall focus on the case (\ref{inter}). Let us check whether the number $v_2^{\pi,ideal}K = 0.32$  is reasonable. For 30-40\% centrality, $v_2\approx 0.07$ according to \cite{Abelev:2008ae}. This is related to $v_2^{ideal}$ as (see Eq.~(61) of \cite{Hatta:2014jva})
\beq
v_2=v_2^{ideal}\left(1-\frac{9}{64}K\right)\,,
\eeq
 which gives $K\approx 2.78$. On the other hand,  $K$ has been independently estimated  in Ref.~\cite{Bhalerao:2005mm}. After correcting the difference in the normalization, we find
$K_{here} \approx 10.7K_{\scriptsize{\mbox{Ref.[13]}}}$, so that $K_{\scriptsize{\mbox{Ref.[13]}}}\approx 0.26$. This is indeed consistent with the range $1/7<K_{\scriptsize{\mbox{Ref.[13]}}}<0.5$  at $\sqrt{s_{NN}}=200$ GeV considered in \cite{Bhalerao:2005mm}.

We now discuss the dependence of $r$ on centrality and energy using the the following formula which can be derived straightforwardly from the results of \cite{Hatta:2015era}
 \beq
r\approx \frac{\Delta v_2^\pi}{A_{ch}} \propto
\frac{\mu_I/T}{A_{ch}} \frac{\eta}{s} \frac{\epsilon}{S^2}\frac{dN}{dy}\,,
\eeq
 where $S$ is the overlapping area of the colliding nuclei and $\epsilon$ is the eccentricity.  We assume that $\eta/s$ is a constant. The part that is sensitive to centrality is then
 \beq
 r\sim \frac{\epsilon}{S^2}\frac{dN}{dy}\,. \label{cen}
 \eeq
  The dependence of (\ref{cen}) on the impact parameter $b$ at $\sqrt{s_{NN}}=200$ GeV can be read off from Table~1 of  \cite{Bhalerao:2005mm}. Converting this $b$-dependence into the centrality-dependence \cite{Broniowski:2001ei}, we obtain the right figure in Fig.~\ref{fig2}. The agreement with the data is remarkably good, except in the most central bin (and the most peripheral bin) where $r$ from the experiment is negative.\footnote{We have seen above that $r$ can become negative if the freezeout temperature is anomalously low. However, this is an unlikely explanation of the sign change because at RHIC the centrality dependence of the  temperature is very weak \cite{Adams:2003xp}. }
 We however note that the systematic uncertainties of the most central and peripheral bins (not included in the error bars in the plot) are very large. In fact, our curve is not inconsistent with the data in these regions if systematic errors are taken into account.

\begin{figure}[h]
   \includegraphics[width=100mm]{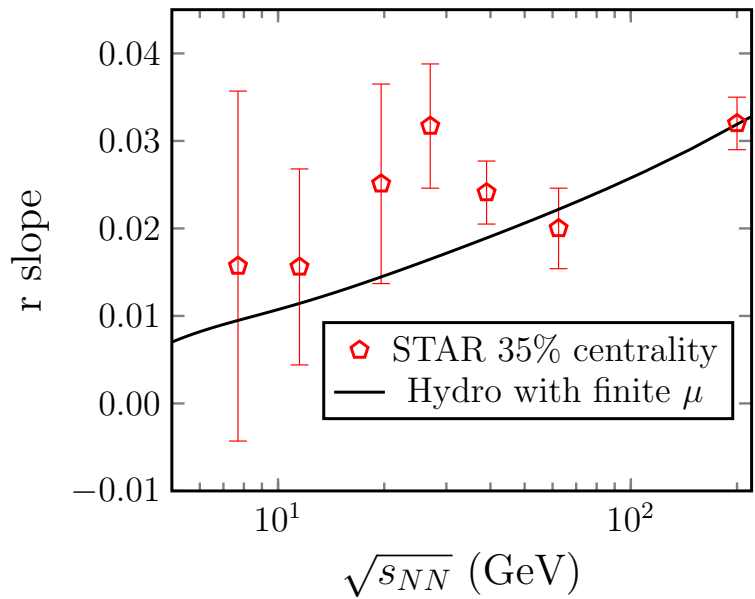}
 \caption{ Slope parameter $r$ as a function of the collision energy at fixed centrality. \label{fig3}}
\end{figure}

On the other hand, the energy dependence of $r$ at fixed centrality is governed by the factor
\beq
r\sim \frac{\mu_I/T}{A_{ch}} \frac{dN}{dy}\,.
\eeq
To evaluate this,  we use the known parametrization $dN/dy \sim (s_{NN})^{0.15}$  \cite{Andronic:2004tx} and compute the ratio $\frac{\mu_I/T}{A_{ch}}$ at each value of the collision energy using the corresponding freezeout parameters  \cite{Stachel:2013zma}. The result is shown in Fig.~\ref{fig3} together with the STAR data. The experimental error bars are rather large especially at low energies, but the general trend is consistent with the theory expectation. We see only a mild energy dependence  since the ratio $\frac{\mu_I/T}{A_{ch}}$ is roughly a constant along the freezeout curve.

In conclusion, we have demonstrated that the STAR result $\Delta v_2^{\pi}\propto A_{ch}$ can be understood within the standard viscous hydrodynamics  without invoking the CMW. Our scenario can be tested in future measurements as follows. The triangular flow difference $\Delta v_3^{\pi}$ satisfies a similar relation to (\ref{our}) and we  find $\Delta v_3^\pi/\Delta v_2^\pi \approx v_3/v_2$. This agrees with the result of \cite{Bzdak:2013yla}, although the mechanisms are different. The kaon $v_2$ difference is dominated by the strangeness chemical potential $\mu_S$: $v_2^{K^-}-v_2^{K^+}\sim -\mu_S$ \cite{Hatta:2015era}. Since $A_{ch}\propto \mu_S$ with a positive proportionality coefficient, the slope $r$ for the kaons should be negative, in contrast to the pion case, and the magnitude is expected to be larger. The situation is similar for the protons.  Note that in these predictions $v_2$ and $v_3$ are integrated over $0<p_T<\infty$. In order to properly test them,  a wider $p_T$ coverage than the one adopted in \cite{Adamczyk:2015eqo} ($0.15\,\mbox{GeV}<p_T<0.5\, \mbox{GeV}$) is necessary:  Higher harmonics $v_n$ are dominated by $p_T\sim nT$  \cite{Hatta:2014jva} and the average $p_T$ increases with hadron mass ($\langle p_T\rangle\sim 0.6$ GeV for kaons at $\sqrt{s_{NN}}=200$ GeV \cite{Abelev:2008ab}).
Preliminary data for these observables can be found in \cite{Shou:2014cja}, but given the narrow coverage in $p_T$ they are not sufficient to draw firm conclusions.

\section*{Acknowledgements}
We thank Jean-Paul Blaizot, Xiao-Feng Luo, Qi-Ye Shou and Nu Xu for discussions and  Anton Andronic for providing the details of the statistical model fits \cite{Stachel:2013zma}.

\end{document}